\documentclass[floatfix,amsmath,aps,prb,eqsecnum]{revtex4-1}
\usepackage{bm}
\usepackage{xcolor}
\usepackage{hyperref}
\usepackage{makeidx}
\usepackage{amsmath}
\usepackage{graphicx}
\usepackage{dcolumn}
\usepackage{amsfonts}
\usepackage{amssymb}
\usepackage[latin9]{inputenc}
\usepackage{color}

\begin{document}

\title{Hybridization induced triplet superconductivity  with $S^z=0$}

\author{Edine Silva$^{1}$}
\author{R. C. Bento Ribeiro$^{1}$}
\author{Heron Caldas$^{2}$}
\author{Mucio A. Continentino$^{1}$}

\email[corresponding author:]{muciocontinentino@gmail.com }
\affiliation{$^{1}$ Centro Brasileiro de Pesquisas F\'{\i}sicas, Rua Dr. Xavier Sigaud, 150, Urca 22290-180, Rio de Janeiro, RJ, Brazil}
\affiliation{$^{2}$ Departamento de Ci\^{e}ncias Naturais, Universidade Federal de S\~{a}o Jo\~{a}o Del Rei, Pra\c{c}a Dom Helv\'{e}cio 74, 36301-160, S\~{a}o Jo\~{a}o Del Rei, MG, Brazil}

\begin{abstract}
The Kitaev superconducting chain is a model of spinless fermions with triplet-like superconductivity. It has raised interest since for some values of its parameters it presents a non-trivial topological  phase that host Majorana fermions. The physical realization of a Kitaev chain is complicated by the scarcity of triplet superconductivity in real physical systems. Many proposals have been put forward to overcome this difficulty and fabricate artificial triplet superconducting chains. In this work we study a superconducting chain of spinful fermions forming Cooper pairs, in a triplet $S=1$ state, but with $S^z=0$.  The motivation is that such pairing can be induced in chains that couple through an antisymmetric hybridization to an s-wave superconducting substrate. We study  the nature of edge states and the topological properties of these chains.  In the presence of a magnetic field the chain can sustain gapless superconductivity with pairs of Fermi points. The momentum space topology of these Fermi points is non-trivial, in the sense that they can only disappear by annihilating each other.  For small magnetic fields, we find well defined degenerate edge modes with finite Zeemann energy.   These modes are not symmetry protected and decay abruptly in the bulk as their energy merges with the continuum of excitations.  

\end{abstract}
\maketitle

\section{Introduction}

The Kitaev chain, a prototypical one-dimensional  toy model of spinless fermions and $p$-wave superconductivity is well known to host Majorana zero modes (MZMs) at the ends of the chain. These spinless fermions form Cooper pairs in a triplet state with $S=1$ and  a finite $S^z$ component of the total spin. The MZMs have, in principle, a prospect of being applied to fault-tolerant topological quantum computation~\cite{Nayak1,Nayak2,Refael}. However,  $p$-wave superconductivity  is very difficult to find in real materials~\cite{Chou,Pan}. One realistic model, proposed to experimentally search for MZMs is the one-dimensional (1D) semiconductor-superconductor (SM-SC) nanowire~\cite{Protocol, para1, para2, v8, v9, tewari2,kotete1,para3}, where these modes are predicted to appear at the ends of the wire under appropriate, specific conditions. The Majorana nanowire consists of a semiconductor with a large Rashba-type SOC in proximity to a conventional s-wave superconductor under a magnetic (Zeeman spin splitting) field to achieve an effective  $p$-wave superconductor~\cite{Pan}.

 In this work we consider  a different model of a $p$-wave superconducting chain. It has spinful fermions~\cite{kotete2,kotfren,bento} that form Cooper pairs in a triplet  state,  $S = 1$, but with $S^z = 0$. 
This type of $p$-wave pairing opens the possibility for obtaining  new chiral superconductors with topological properties~\cite{Robinson,Amitava}.  

The main motivation for studying these chains is the following. When  atoms are deposited on a superconducting substrate, in general its orbitals hybridize with those of the substrate~\cite{newns}  providing a mechanism for inducing superconductivity in the chain.  If the  substrate is a BCS superconductor and the hybridization $V_{ij}$ between the orbitals at site $i$ of the substrate and at site $j$ of the chain is  antisymmetric, i.e., $V_{ji}=-V_{ij}$, or in momentum space $V_{-k}=-V_k$, the {\it induced superconductivity} in the chain is of the type $S=1$, $S^z=0$. The energy scale involved in this process can be significant~\cite{newns}.

The mechanism that induces this type of superconductivity in the wire is similar to that occurring  in multi-band superconductors~\cite{Annica,Annica2,Filipe,Fernanda,bianconi,new,new2}. Consider a two-band ($a$ and $b$)  superconductor with an attractive interaction in  the $a$ band that gives rise to a BCS pairing of the electrons in this band (substrate). An anti-symmetric hybridization $V_k$ between  these electrons  and those in the non-interacting $b$-band  induces a pairing gap with $p$-wave symmetry, of the type $S=1$, $S^z=0$ in the $b$-band (the chain)~\cite{Fernanda,Filipe,Annica}. The induced $p$-wave pairing $\Delta_{ind}^p(k)$, with $S=1$, $S^z=0$, is given by~\cite{Fernanda,Filipe,Annica,Annica2}, (see also note~\cite{note}),
\begin{eqnarray}
\Delta_{ind}^p(k)=  \frac{ V_k}{\sqrt{(\epsilon_k^b-\epsilon_k^a)^2 + 4 |V_k|^2}}\Delta_a^s,
\label{eq1}
\end{eqnarray}
where the quasi-particles in the interacting  $a$-band, with a dispersion relation $\epsilon_k^a$ in the normal state, condense in an s-wave BCS singlet superconducting state with an $s$-wave gap $\Delta_s^a$.  The dispersion of the non-interacting band is given by, $\epsilon_k^b$. The anti-symmetric hybridization $V_k$ between the orbitals of different parities in the interacting and non-interacting bands is responsible for an induced  $p$-wave pairing  $\Delta_{ind}^p(k)$, of the type $S=1$, $S^z=0$,  as  given by Eq.~\ref{eq1}(see note~\cite{note}). Notice that  $V_k$ in Eq.~\ref{eq1} is a {\it one-body term}, or simply,  an inter-band hopping that transfer electrons between orbitals of different parities between the  bands. This mechanism for induced superconductivity is different from the usual proximity effect that involves Cooper pair tunneling or Andreev reflections at the non-superconductor-superconductor boundary~\cite{Stanescu,Loo}.

In this paper we consider a model that describes a BCS superconducting substrate on top of which is deposited a chain with non-interacting electrons. This kind of setup has already been implemented experimentally~\cite{new2,apora}. The electrons in the chain,   the non-interacting $b$-band,  hybridize with those in the $a$-band of the superconducting substrate where the Cooper pairs are formed.  If this hybridization is anti-symmetric,  the induced superconductivity in the chain is unconventional and corresponds to the pairing $\Delta_{ind}^p$ in Eq.~\ref{eq1}.   In this equation, $\Delta_a^s$ is the BCS $s$-wave pairing of the substrate~\cite{note}.  The model is valid whenever the main coupling between the chain and substrate is through the hybridization between their orbitals. Notice that the hybridization, contrary to the spin-orbit interaction, has no spin-flip terms and the Cooper pairs in the chains preserve the anti-parallel coupling of the spins inherited from the BCS Cooper pairs of the substrate. The antisymmetric character of the total wave-function of the induced Cooper pair is conferred  by the antisymmetric hybridization $V_k$  that plays the role of the spatial dependent wave function of the pair  and  allows for a symmetric state of the spins~\cite{note}.  Then, the superconductivity induced in the chain is triplet  $S=1$, but with $S^z=0$.
Notice that the antisymmetric character of  the  hybridization arises when it mixes orbitals with angular momenta that differ by an odd number~\cite{Fernanda,Filipe,Annica}. This  includes the important cases that the orbitals in the chain-substrate system have $s-p$ or $p-d$ character. 
For completeness the model also includes the spin-orbit coupling (SOC) in the wire and the effect of a magnetic field. In the case of SOC, this limits the validity of our  results to the case that the spin-orbit interaction vanishes or is smaller than the induced superconducting parameter in the wire.

With the motivation of a physical mechanism to obtain an $S=1$, $S^z=0$, $p$-wave-superconducting chain, and its possible applications~\cite{superspin}, we present here a study of the topological properties and excitations  of such a chain. In the absence of a magnetic field and the conditions stated above for the SOC, the $S=1$, $S^z=0$ chain is topological, for a range of parameters. It presents four Majorana modes, two in each extremity of a finite chain. As the magnetic field is turned on, these modes acquire a finite Zeeman energy. They are not symmetry protected and disappear as  they merge with the continuum of Bogoliubov excitations for a sufficiently high magnetic field. 

The $S=1$, $S^z=0$ chain,  in the presence of a large magnetic field~\cite{kotete1}, presents gapless superconductivity due to the presence of pairs of Fermi points~\cite{volovik,3dSOC}.  The non-trivial momentum space topology of these Fermi points implies their stability, since they can only be destroyed by annihilating in pairs~\cite{horava,voz}. The dispersion at the Fermi points is linear, like in Dirac points. This is in contrast with the Kitaev chain that is a gapped superconductor with  Majorana edge modes. 

The paper is organized as follows. In Section II we present the Hamiltonian of the model. In Section III we obtain the energy spectra of an infinite chain. In Section IV we study a finite chain described by the Hamiltonian written in terms of Majorana operators. The numerical results are shown in Section V. Section VI is devoted to the calculation  of the topological invariant and  topological indexes of the several phases of the model. We conclude in Section VII.

\section{Hamiltonian}

The Hamiltonian describing the  chain of N sites and spinful  fermions with a $S=1$, $S^z=0$, $p$-wave  induced superconducting interaction is given by $\mathcal{H}=\mathcal{H_N}+\mathcal{H_{HS}}$, where

\begin{eqnarray}
\label{eq2}
\mathcal{H_N}& =& -\mu \sum_{ j =1, \sigma}^N  c^{\dagger}_{ j, \sigma}c_{ j, \sigma}- h \sum_{ j =1, \sigma}^N  \sigma c^{\dagger}_{ j, \sigma}c_{ j, \sigma}-  \sum_{  j=1, \sigma}^{N-1}  \left( t \thinspace c^{\dagger}_{ j , \sigma} c_{  j+1, \sigma}+h.c\right)+ i \lambda \sum_{j=1,\sigma,\bar{\sigma}}^{N-1} \left(c^{\dagger}_{ j, \sigma}(\sigma^y)_{\sigma \bar{\sigma}} c_{ j+1,  \bar{\sigma} }+h.c. \right)
\nonumber \\ 
\mathcal{H_{HS}}&=& -\frac{1}{2} \sum_{ j=1, \sigma}^{N-1}\left( \Delta (c^{\dagger}_{ j, \sigma}c^{\dagger}_{ j+1,  -\sigma} - c^{\dagger}_{ j+1, \sigma }c^{\dagger}_{j, -\sigma }) + h.c.  \right).
\end{eqnarray}
The first equation describes the normal chain in the presence of a uniform external magnetic field $h$ parallel to the wire~\cite{para1,para2,para3}. The quantity $\mu$  is the chemical potential, $t$ a nearest neighbor hopping. In our strict one-dimensional model, the Rashba-like term   is essentially an antisymmetric spin-flip hopping $\lambda$ due to the spin-orbit interaction~\cite{Protocol,Pan,tewari2,lawz,law}.

The Hamiltonian $\mathcal{H_{HS}}$ represents the induced superconductivity in the chain due to its hybridization with the BCS superconducting substrate (see Eq.~\ref{eq1}). Since this hybridization is antisymmetric, $\Delta$ is the induced antisymmetric superconducting pairing ($\Delta_{ij}=-\Delta_{ji}=\Delta$)  between fermions with antiparallel spins in neighboring sites of the wire.  In momentum space it is defined by Eq.~\ref{eq1}, where we removed all indices for simplicity.  N is the number of sites in the chain, and $\sigma = \pm$, corresponds to spin up and spin down, respectively. The total antisymmetry  of the order parameter with  antiparallel spins is guaranteed by the spatial antisymmetric wave function of the Cooper pairs.  Then the induced superconductivity is a triplet state with $S=1$, but with the  z-component of the total spin of the Cooper pair, $S^z=0$. The term $h.c.$ stands for Hermitian conjugate.

\section{Infinite chain}

For an infinite chain with periodic boundary conditions we can Fourier transform   Hamiltonian Eq.~(\ref{eq1}) in momentum space. We choose the basis~$\Psi_{{\bf k}} = (c_{\textbf{k},\sigma}^{\dag}, c_{\textbf{k}, -\sigma}^{\dag},
c_{-\textbf{k},\sigma}, c_{\textbf{-k},-\sigma})^{T}$, with $\sigma = \uparrow$ and $-\sigma=\downarrow$ to obtain,

\begin{eqnarray}
H=\frac{1}{2}\sum_{{\bf k}}\Psi_{{\bf k}}^{\dag}\mathcal{H}({\bf k})\Psi_{{\bf k}}
 + \frac{1}{2} \sum_{\textbf{k}} [\varepsilon_{{\bf k} \uparrow} + \varepsilon_{-{\bf k} \downarrow} ], 
\label{6}
\end{eqnarray}
where $\varepsilon_{{\bf k} \uparrow \downarrow} = -(2 t \cos k + \mu \pm h) $, $\lambda_k= 2 i \lambda \sin(k)$ and $\Delta_k=  2 i \Delta \sin k$,
\begin{eqnarray}
\mathcal{H}({k})=\left(
  \begin{array}{cccc}
    \varepsilon_{{\bf k} \sigma} & \lambda_{k}  & 0 & \Delta^*_k  \\
\lambda^*_{k}  & \varepsilon_{{\bf k} -\sigma} & \Delta^*_k & 0\\
   0& \Delta_k &  - \varepsilon_{-{\bf k} \sigma} &  -\lambda^*_{-k} \\
   \Delta_k & 0& -\lambda_{-k} &  -\varepsilon_{-{\bf k} -\sigma}  \\
  \end{array}
\right).
\label{7}
\end{eqnarray}
Notice that $\lambda_{-k}^*=\lambda_k$, with a similar relation for $\Delta_k$. We remark that in the present Hamiltonian, the gap parameter is $k$-dependent and has p-wave symmetry, distinctively from other previous approaches, as for instance, in Refs.~\cite{v8,3dSOC,para2}.

\begin{figure}[ht]
\centering
\includegraphics[width=0.4\columnwidth]{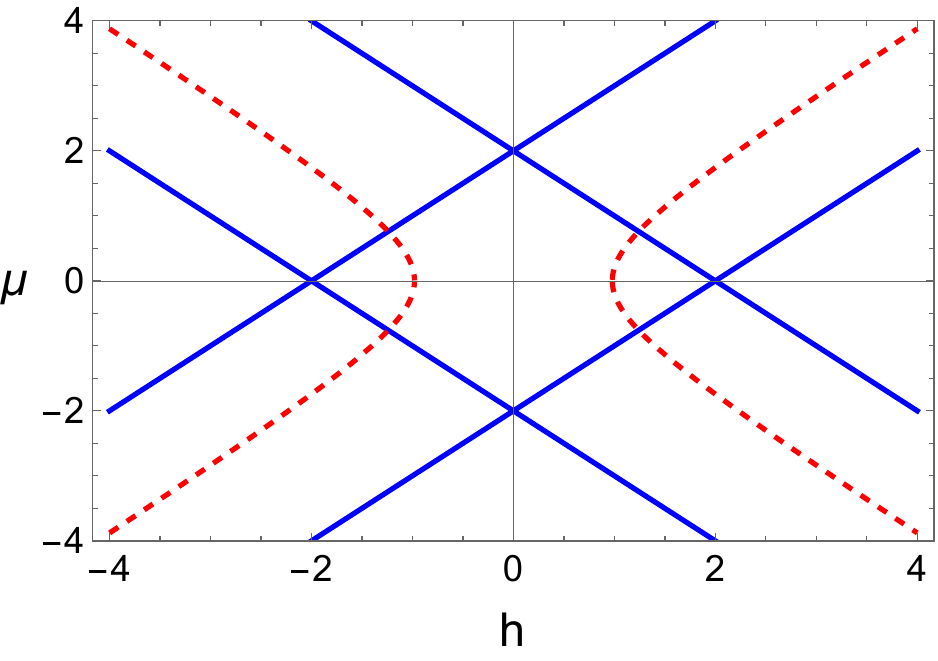}
\caption{(Color online) Gap closing lines. At the blue lines, $h_=\pm2t \pm \mu$ the gap closes at the time-reversal invariant momenta $k=0$ and $k=\pm \pi$. Along the red (dashed) lines, the gap closes at   $k=\pm \pi/2$ (we consider only the case, $\lambda <\Delta$). In the figure we chose $t$ as the unity of energy ($t=1$). The red dashed lines  correspond to $\lambda =0.1$ and $\Delta=0.5$. }
\label{fig1}
\end{figure}

The Hamiltonian, Eq.~\ref{7}, can be written as $\mathcal{H}=(\epsilon_k-\mu)\sigma_z \otimes \tau_0 - h \sigma_z \otimes \tau_z + i \lambda_k \sigma_z \otimes \tau_y -i \Delta_k \sigma_y \otimes \tau_x$, where $\sigma_i$ and $\tau_i$ are Pauli matrices ($i=x, y, z$) and $\tau_0=\sigma_0$ are the 2 x 2  identity matrix and $\epsilon_k=-2t \cos k$. It has particle-hole symmetry since $U_p \mathcal{H^*} (k)U_p^{\dagger}= - \mathcal{H}(-k)$, where $U_p=i \sigma_x \otimes \sigma_0$. On the other hand, we have, $U_t  \mathcal{H^*}(k) U_t^{\dagger} \ne \mathcal{H}(-k)$, with $U_t=i \sigma_0 \otimes \sigma_y$,  implying that the system is not time-reversal invariant. Time-reversal invariance is broken by the magnetic field, but also by the triplet pairing of the quasi-particles. 
The Hamiltonian, Eq.~\ref{7}, has an additional chiral symmetry that will be discussed further on.

The Hamiltonian of the infinite chain can be diagonalized and the dispersion relations of the quasi-particles are given by, 
\begin{eqnarray}
\omega_{1}(k)&=&  \sqrt{\epsilon_k^2 + h^2 + \lambda_k^2+\Delta_k^2 + 2 \sqrt{(h^2+\lambda_k^2)(\epsilon_k^2+\Delta_k^2 )} } \nonumber \\
\omega_{2}(k)&=&  \sqrt{\epsilon_k^2 + h^2 + \lambda_k^2+\Delta_k^2 -  2 \sqrt{(h^2+\lambda_k^2)(\epsilon_k^2+\Delta_k^2 )} },
\end{eqnarray}
$\omega_{3}(k)=-\omega_{1}(k)$ and $\omega_{4}(k)=-\omega_{2}(k)$. 
These dispersions have several {\it gap closing lines}, as shown in Fig.\ref{fig1}.  Closing of the gaps occur at the time-reversal invariant points of the Brillouin zone, $k=0$ and $k=\pm \pi$, but also for $k=\pm \pi/2$. In the former case the critical fields at which the gap closes are given by, $h_c=\pm 2t \pm \mu$ and are independent of the other parameters of the model. 

The critical fields for gap closing at $k\pm \pi/2$ are given by, $h_c=\pm \sqrt{4 (\Delta^2- \lambda^2)+\mu^2}$ and depend on $\Delta$ and $\lambda$, besides the chemical potential $\mu$.   We can  distinguish two cases, $\lambda>\Delta$ and $\lambda<\Delta$, but here we consider only the case of small spin orbit coupling, $\lambda<\Delta$.
Notice that for $\mu=0$, and assuming $\lambda=0$, for simplicity, the first gap closing with increasing field occurs for $k_{\pm}=\pm \pi/2$ and $h_c=\pm  2\Delta$ ($\Delta<2t$). This vanishing of the gap at a wavevector that is not time-reversal invariant is particularly interesting and we will explore this case numerically further on. We will also discuss the topological nature of the different phases of the model.

\section{Finite size chains}

For the purpose of studying the edge modes in the system, we  consider a finite size chain with $N$ sites and open boundary conditions.  We neglect spin-orbit coupling ($\lambda=0$), for simplicity, in this section. 
We rewrite Eq.~\ref{eq1} in terms of new real operators
\begin{eqnarray}
c_{j,  \sigma }=\frac{1}{2}(\gamma^{B}_{ j, \sigma} + i \gamma^{A}_{j,  \sigma} )\\ \nonumber
c^{\dagger}_{j,  \sigma }=\frac{1}{2}(\gamma^{B}_{ j, \sigma} - i \gamma^{A}_{j,  \sigma} )
\end{eqnarray}
where the Majorana operators satisfy, $\gamma^{\beta}_ {j. \sigma } =\gamma^{\beta \dagger}_ {j, \sigma }$,  $\{\gamma^{\beta }_ {i, \sigma }, \gamma^{\beta}_ {j, \sigma^{\prime} }\}=\delta_{ij,\sigma \sigma^{\prime}}$ and $\gamma^{\beta}_ {j. \sigma }\gamma^{\beta}_ {j. \sigma }=1$ ($\beta=A,B$). 
The symbol $\sigma$ in  $\gamma^{B}_{ j, \sigma}$ {\it does not mean a Majorana of spin $\sigma$}. It is only a label to distinguish the various operators, since we need two Majoranas to represent an electron of spin up and two for an electron of spin down.

\begin{figure}[ht]
\centering
\includegraphics[width=0.5\columnwidth]{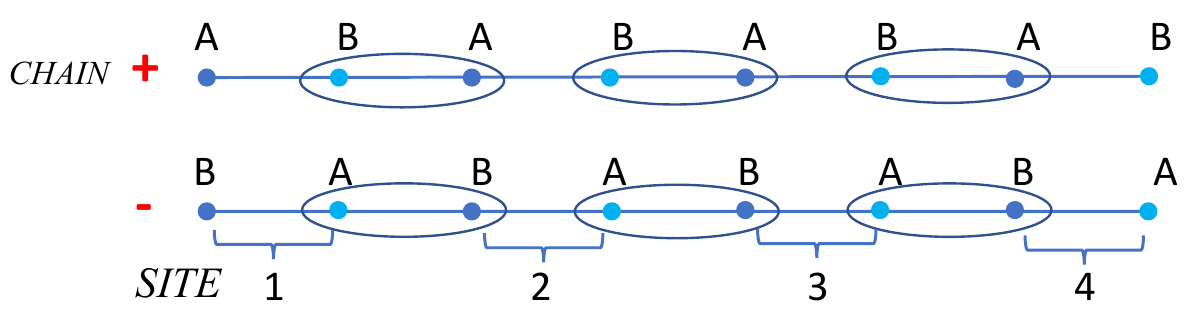}
\caption{(Color online) Eq.~\ref{Hmaj} describes two independent sub-chains ($\pm$) in terms of the operators $\alpha^{\pm}$. The figure represents a topological phase, with Majoranas at the edges of each sub-chain.}
\label{fig2}
\end{figure}

In terms of these new operators the Hamiltonian can be rewritten as,
\begin{eqnarray}
\mathcal{H}&=&-\frac{1}{2}\sum_{j=1, \sigma}^N (\mu +\sigma h)(1+ i \gamma^B_{j, \sigma } \gamma^A_{j, \sigma })-\frac{i t}{4} \sum_{j=1, \sigma}^{N-1} (\gamma^B_ {j, \sigma } \gamma^A_ { j+1, \sigma}-\gamma^A_ {j, \sigma } \gamma^B_ { j+1, \sigma}) - \\ \nonumber
&&\frac{i \Delta}{4} \sum_{j=1, \sigma}^{N-1} (\gamma^B_ {j, \sigma } \gamma^A_ { j+1, -\sigma}+\gamma^A_ {j, \sigma} \gamma^B_ { j+1, -\sigma}).
\end{eqnarray}
Next, we introduce more four Majorana operators given by,
\begin{eqnarray}
\alpha^{A \pm}_{j, \sigma }= \gamma^A_{j, \sigma} \pm \gamma^A_{j, -\sigma }\\ \nonumber
\alpha^{B \pm}_{j, \sigma }= \gamma^B_{j, \sigma } \pm \gamma^B_{j, -\sigma }.
\end{eqnarray}
Notice that 
\begin{eqnarray}
\label{qrt}
\alpha^{A \pm}_{j,  \sigma }=\pm \alpha^{A \pm}_{j, -\sigma } \\ \nonumber 
\alpha^{B \pm}_{j,  \sigma }=\pm \alpha^{B \pm}_{j, -\sigma }.
\end{eqnarray}
Finally, in term of these new operators, the Hamiltonian of the superconducting chain can be written as,
\begin{eqnarray}
\label{basic}
\mathcal{H}=-\frac{i \mu}{4} \sum_{j=1}^{N} \left(\alpha^{B +}_{j, \uparrow} \alpha^{A +}_{j, \uparrow} + \alpha^{B -}_{j, \uparrow} \alpha^{A -}_{j, \uparrow} \right)  
-\frac{i}{8} \sum_{j=1}^{N-1}  (t+\Delta)\left(\alpha^{B +}_{j,  \uparrow} \alpha^{A +}_{j+1,  \uparrow} - \alpha^{A -}_{j,  \uparrow} \alpha^{B -}_{j+1,  \uparrow} \right) \\ \nonumber
-\frac{i}{8} \sum_{j=1}^{N-1} (t-\Delta)\left(\alpha^{B -}_{j,  \uparrow} \alpha^{A -}_{j+1,  \uparrow} - \alpha^{A +}_{j,  \uparrow} \alpha^{B +}_{j+1,  \uparrow} \right) 
-\frac{i h}{4} \sum_{j=1}^{N} \left(\alpha^{B +}_{j,  \uparrow} \alpha^{A -}_{j,  \uparrow} + \alpha^{B -}_{j,  \uparrow} \alpha^{A +}_{j, \uparrow} \right),
\end{eqnarray}
where the sum over $\sigma$ has been performed and expressed in terms of $\sigma =\uparrow$. This index  $\sigma =\uparrow$ now becomes redundant but we keep it anyway. The $\alpha$-operators are such that,
\begin{eqnarray}
\alpha^{A \pm}_{\ell,  \uparrow }= \gamma^A_{\ell, \uparrow} \pm \gamma^A_{\ell, \downarrow }\\ \nonumber
\alpha^{B \pm}_{\ell,  \uparrow }= \gamma^B_{\ell, \uparrow } \pm \gamma^B_{\ell, \downarrow }.
\end{eqnarray}
The Hamiltonian Eq.~\ref{basic} describes two independent ($\pm$) sub-chains that are coupled by the magnetic field term, the last term in Eq.~\ref{basic} (see Fig~\ref{fig2}).  
Notice from Eqs.~\ref{qrt} and Eq.~\ref{basic} that the magnetic field term breaks time-reversal symmetry, as it is not invariant under the change $\sigma \rightarrow - \sigma$.

For $t=\Delta=0$, the Hamiltonian, Eq.~\ref{basic},  describes  trivial chains  coupled by the magnetic field. On the other hand, if we take $\mu=h=0$  and $t=\Delta$, we obtain,
\begin{equation}
\label{Hmaj}
\mathcal{H}=\frac{-i \Delta}{4} \sum_{j=1}^{N-1}\left(\alpha^{B +}_{j, \uparrow} \alpha^{A +}_{j+1, \uparrow} - \alpha^{A -}_{j, \uparrow} \alpha^{B -}_{j+1, \uparrow} \right).
\end{equation}
Now the system consists of two decoupled $\pm$ chains, as shown  schematically in Fig.~\ref{fig2}. The Majoranas $\alpha^{B -}_{1, \uparrow}$ and $\alpha^{A +}_{1, \uparrow}$ at the beginning of these chains do not enter the Hamiltonian, as also the Majoranas $\alpha^{B +}_{N, \uparrow}$ and $\alpha^{A -}_{N, \uparrow}$ at the ends of the chains. These are the zero energy edge modes that signal the existence of a non-trivial topological phase in the superconducting chain. Notice that the $\pm$ chains are {\it not} associated with a given spin direction. As we show below, these zero modes persist for $|\mu/2t| <1$, which characterizes the topological phase of the system in the absence of a magnetic field ($h=0$).

Then, for $h=\mu=0$  and $t=\Delta$, the system is formed of two independent Kitaev-like chains, each with two Majoranas. We can combine the Majoranas at the edges of each chain to obtain, 
\begin{eqnarray}
g^-=\alpha^{B -}_{1, \uparrow}+i \alpha^{A -}_{N, \uparrow} \\ \nonumber
g^+=\alpha^{B +}_{N, \uparrow}+i \alpha^{A +}_{1, \uparrow},
\end{eqnarray}
to form a pair of non-local fermions, one in each sub-chain.  Then, in the absence of a magnetic field, which couples the $\pm$ sub-chains, the latter are the zero energy edge modes of the system. In terms of the original fermions operators, we have
\begin{equation}
g^-=(c_{1, \uparrow }-c^{\dagger}_{N,  \uparrow})- (c^{\dagger}_{1, \downarrow}+c_{N, \downarrow})
-\left( (c_{1, \downarrow}-c^{\dagger}_{N, \downarrow})-(c^{\dagger}_{1, \uparrow }+c_{N, \uparrow}) \right).
\end{equation}
We recall that in the Kitaev spinless p-wave superconducting chain, with $S=1$, $S^z=1$, the non-local fermion is given by~\cite{alicea},  
\begin{equation}
f_K=(c_1-c^{\dagger}_{N}) -(c^{\dagger}_{1}+c_{N}).
\end{equation}

In the presence of a magnetic field, we can combine two Majoranas, either in the same sites or on different edges, to form the fermions. In the former case we have,
\begin{eqnarray}
g_1&=&\alpha^{B -}_{1, \uparrow}+i \alpha^{A +}_{1, \uparrow} \\ \nonumber
g_N&=&\alpha^{B +}_{N, \uparrow}+i \alpha^{A -}_{N, \uparrow},
\end{eqnarray}
or in terms of fermion operators,
\begin{eqnarray}
g_1&=&2(c_{1, \uparrow}-c^{\dagger}_{1, \downarrow}) \\ \nonumber
g_N&=&2(c_{N, \uparrow}+c^{\dagger}_{N, \downarrow}).
\end{eqnarray}
These modes are present for $|\mu/2t| <1$, for $0<h<h^*$, i.e., for finite fields but  below a critical field $h_c$, as we show below. They are localized at the edges of the chain and have a finite  Zeemann energy. They are not symmetry protected,  as they exist in a region of the phase diagram  that is topologically trivial, as evidenced by the topological invariant calculated further on.

\section{Numerical results}

In this section, we discuss the dispersion relations for an infinite chain and obtain numerical results on finite chains of size $N$ with open boundary conditions, in the presence of SOC and magnetic field.
We consider two different situations that are distinguished by the points of the Brillouin zone at which the gap closes. This may occur either at the time-reversal invariant momenta, $k=0$, $k = \pm \pi$, or at the points $k=\pm \pi/2$. 

\begin{figure}[ht]
\centering
\includegraphics[width=0.35\columnwidth]{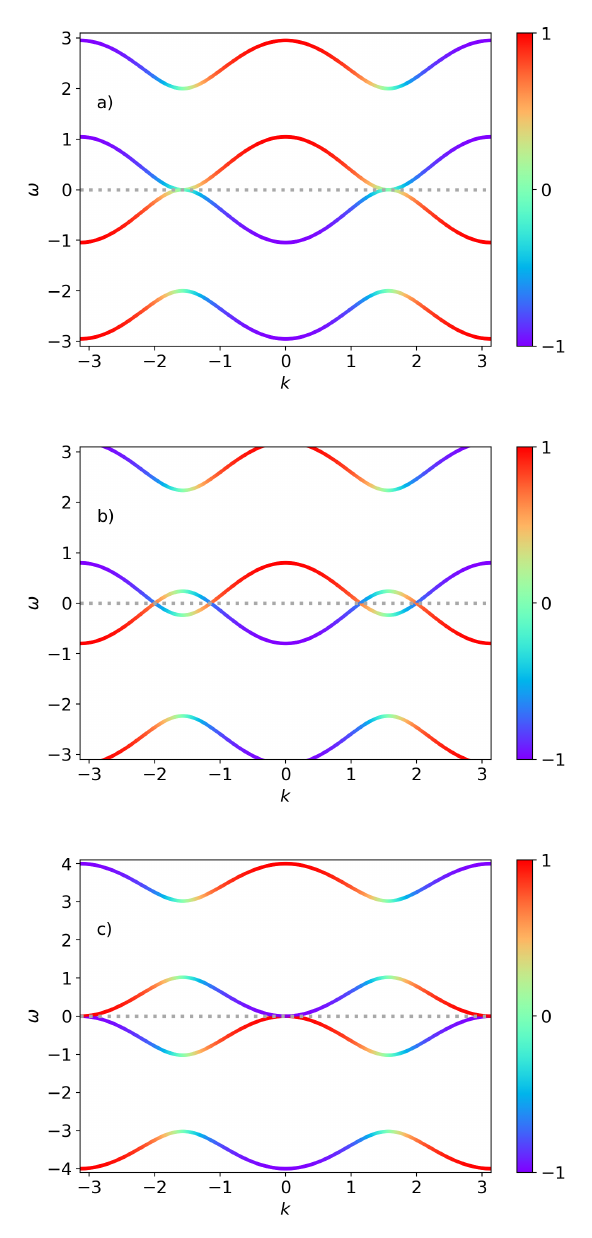}
\caption{(Color online) a) Dispersion relations for $t=1$, $\mu=0$, $\lambda=0.15$ and $\Delta=0.5$  a)   gap closing at $k=\pi/2$ for the critical field $h_c=\sqrt{4(\Delta^2-\lambda^2)}$. b) Two pairs of monopoles in the non-trivial topological phase for  $h_c <h <2t$ ($h=1.2$). c)  For $h=2t$ the two pairs of monopoles  annihilate at $k=0$, and $k=\pm \pi$). The color code indicates the degree of superposition between particles and hole states.}
\label{fig3}
\end{figure}

\subsection{Gap closing at $k= \pm  \pi/2$}

First, for the infinite chain, with $\mu=\lambda=0$, the superconducting gap decreases with increasing magnetic field and finally closes at the non-time-reversal wave-vectors, $k=\pm \pi/2$ at a critical magnetic field $h_c=  2 \Delta$, as shown in the dispersion relations of Fig.~\ref{fig3}. For magnetic fields $2\Delta <h <2t$, the system enters a topological phase characterized by the presence of pairs of monopoles ~\cite{3dSOC,volovik,murakami} at the field-dependent wave-vectors $k =\pm k_1$ and $k=\pm k_2$, as shown in Fig.~\ref{fig3}b. Finally, for $h=2t$ the monopoles annihilate each other at the time-reversal k-points of the Brillouin zone, Fig.~\ref{fig3}c. These results are not qualitatively affected by the presence of the spin-orbit interaction, as can be seen in Fig.~\ref{fig3}, where it is taken finite. In the next section we will give a detailed analysis of the topological nature of this phase.
\begin{figure}[ht]
\centering
\includegraphics[width=0.4\columnwidth]{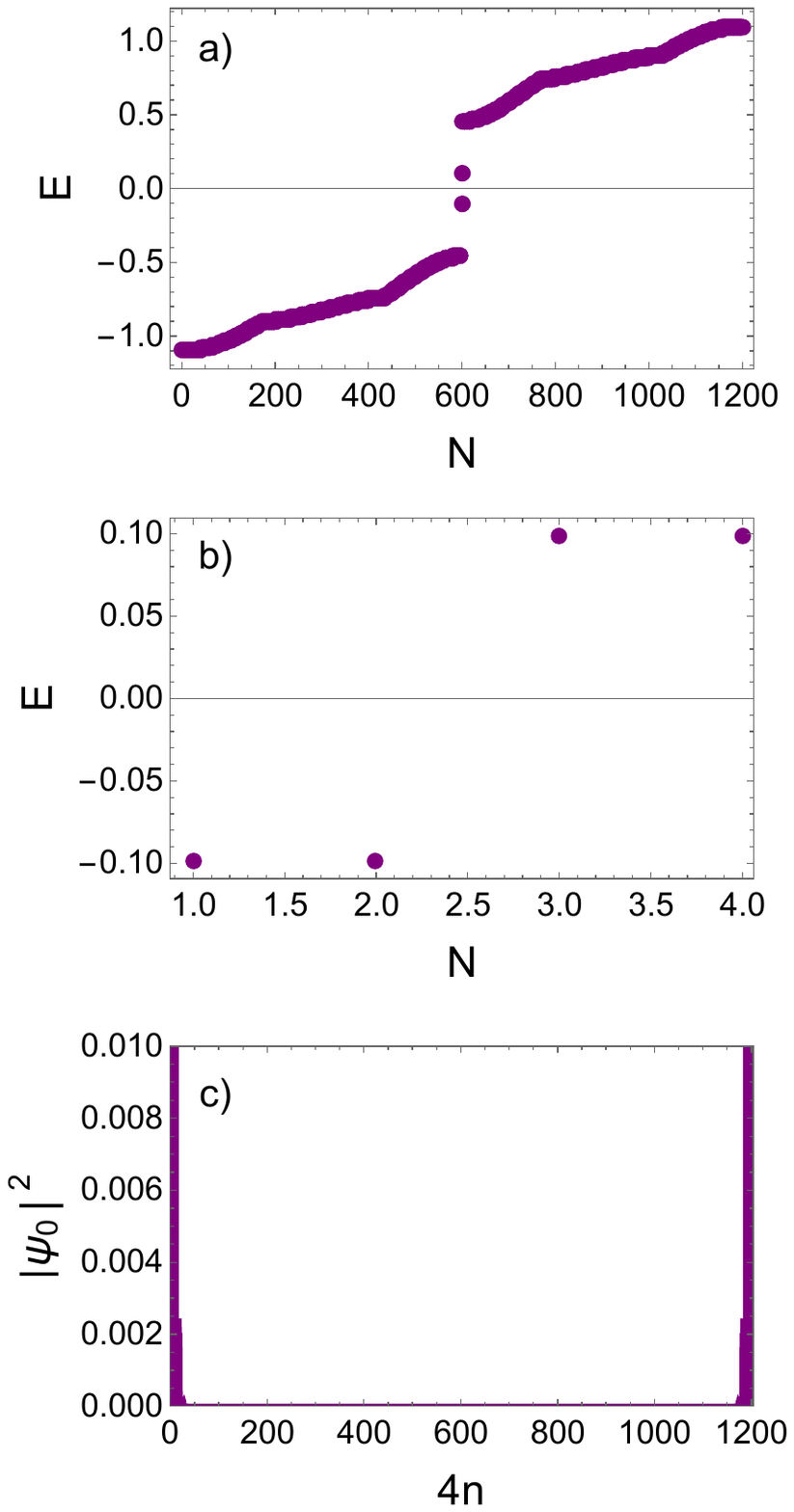}
\caption{(Color online) Spectrum of eigenvalues for a chain of 300 sites with $\mu=0$, $\Delta=0.6$, $\lambda=0.1$ and $h=0.2$, in units of the hopping $t$ ($t=1$). In b) the energy of the 4 edge modes, with two degenerate, is emphasized. In c)  the wave functions of the  modes shown in b (see also Fig.~\ref{fig5}).}
\label{fig4}
\end{figure}
\begin{figure}[ht]
\centering
\includegraphics[width=0.4\columnwidth]{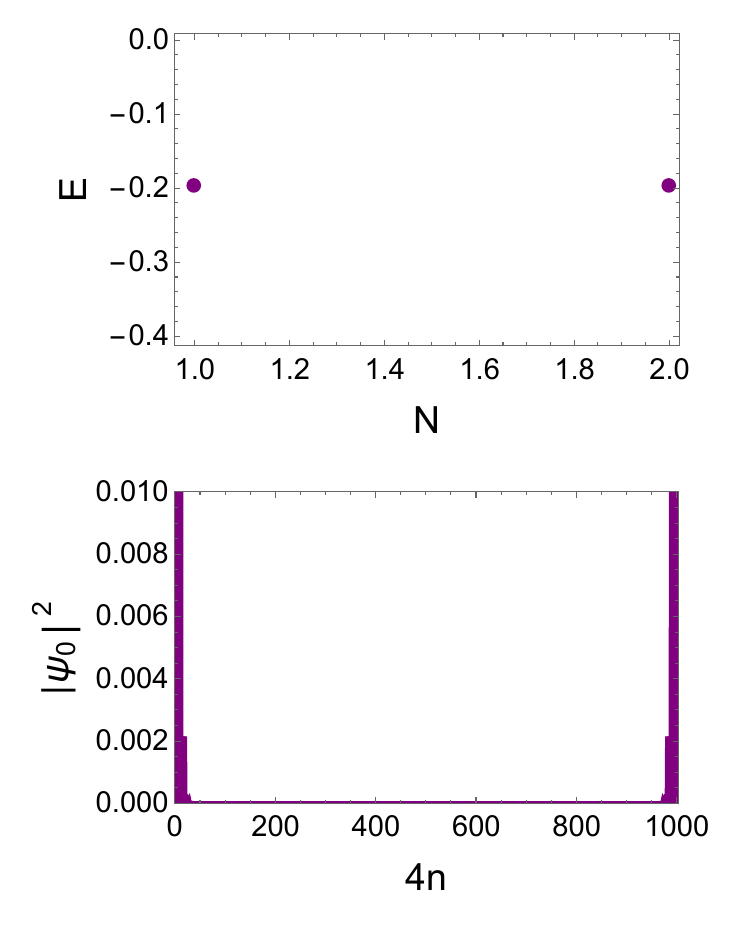}
\caption{(Color online) The energy of the edge modes aligned with the magnetic field and their wave functions, for $\mu=0$, $\Delta=0.6$, $\lambda=0.1$ and $h=0.4$. These modes combine to a form a single quasi-particle with the correct total Zeemann energy. This quasi-particle is delocalized since its wave function resides  in the two opposite edges of the chain. }
\label{fig5}
\end{figure}
Notice that we  are neglecting the possibility of any finite momentum pairing, since the superconductivity in the wire is induced by the   s-wave substrate.     
\begin{figure}[ht]
\centering
\includegraphics[width=0.35\columnwidth]{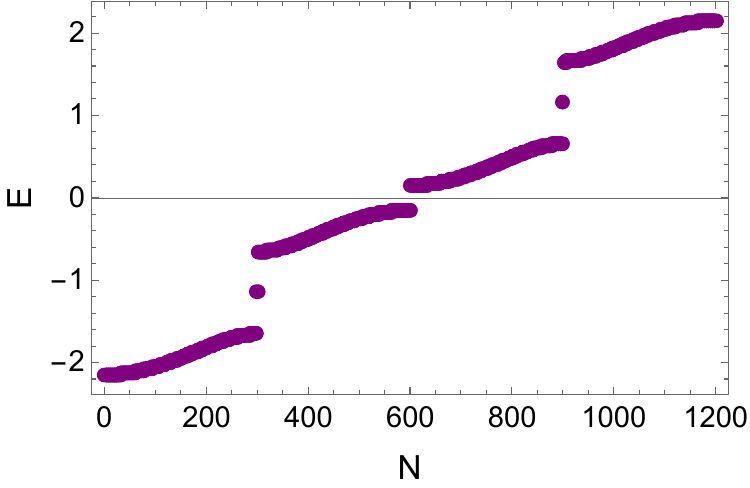}
\caption{(Color online) As a new gap opens and the Zeemann energy falls in a gap, the edge modes reappear. In this case at the trivial topological phase with $\mathcal{M}=0$, for $t=1$, $h=2.3$, $\Delta=0.5$, $\lambda=0.1$ and $\mu=0$ (see Fig.~\ref{fig7}).}
\label{fig6}
\end{figure}

For finite chains at $h=0$, we confirm numerically, the presence of four zero energy Majorana modes, two at each edge of the chain, for $|\mu/2t| \le 1$. Their wave functions are localized at the ends of the chain. This is  expected since in this limit of $h=0$, the system consists of two decoupled Kitaev chains, as we obtained  before. Then, for $h=0$ we find a topological phase with zero energy edge modes for  $|\mu/2t|<1$. 

As we turn on the magnetic field in the finite chain, the four edge modes persist in the presence of a small  field, but now they are separated in two double degenerate modes with finite energies, as shown in Fig.~\ref{fig4}. The lower energy states correspond to the Zeemann energy of an electron with spin parallel to the field and the excited state, with positive energy,  to  that of an electron with spin antiparallel to the field. The wave functions of these edge modes decay in the bulk with a penetration length which is nearly field independent and corresponds to that of the system in the absence of the field. Further increasing the magnetic field ($\mu=\lambda=0$), as it reaches the value, $h^{*}=\Delta$, i.e., {\it before the gap closes}, the energy of the local modes merge with the continuum of Bogoliubon excitations and their wave functions become abruptly delocalized. The closest analogy of this phenomenon is the Chandrasekhar-Clogston limit in conventional superconductors~\cite{cc}  where the magnetic field is screened up to a critical field beyond which it penetrates abruptly in the bulk. 

Further increasing the field, the gap decreases and  closes for a critical field $h_c=\pm \sqrt{4 (\Delta^2- \lambda^2)+\mu^2}$, in the presence of SOC ($\lambda<\Delta$).  The effect of SOC in this case is to compete with the superconducting  gap, reducing it.  Finally, for $h>h_c$, Fermi points appear in the system. These Fermi points are topologically protected as we discuss further on.
Notice  in Fig.~\ref{fig4}a  that the spectra are  always particle-hole symmetric and this holds for any values of the parameters of the Hamiltonian, in agreement with what was obtained previously. Further on, we will discuss the existence  of a hidden chiral symmetry of the Hamiltonian.

It is interesting to investigate how the edge modes  give rise to a full electronic spin. They can  combine either on the same edge or on different edges as a delocalized quasi-particle.   
The numerical results in Fig.~\ref{fig5} show that the wave functions of the modes with negative energy, i.e., with spins aligned parallel to the field are localized on {\it two opposite edges of the chain}. Consequently, the wave function of the {\it electron}  with spin up,  parallel  to the magnetic field, is delocalized with equal weight on different edges of the chain, as shown in Fig.~\ref{fig5}. The same holds for the excited state with spin down. Notice also in Fig.~\ref{fig5} that the energy of two edge modes are required to give the Zeeman energy of a single electron.
  
Finally it is shown  in Fig.~\ref{fig6},  that for large enough fields,  local modes reappear when their Zeemann energy falls in a gap of the spectra of bogoliubons.

\subsection{Gap closing at time-reversal wave vectors}

The gap closing phenomena that occurs at the time-reversal invariant  wave-vectors in the present study are  associated with the annihilation of  monopoles, shown  in Fig.~\ref{fig3}c. Monopoles of different charges annihilate each other exactly at these points, $k=0$ and $k=\pm \pi$.  These monopoles, or Fermi points are non-trivial topological objects in momentum space~\cite{volovik}. They are characterized by a topological charge, given by a winding number~\cite{volovik}.

For $h=0$ we observe gap closing as a function of the chemical potential for $\mu=\pm 2t$ at the time-reversal point $k=0$. Also for $h=0$ the spin-orbit interaction can  promote gap closing for $\lambda=\Delta$. However, we restrict our study here for the case $\lambda < \Delta$.

\section{Topological invariant}

In order to characterize the different phases and transitions at the gap closing points, we calculate  topological indices for our problem. First we seek for a chiral symmetry of the Hamiltonian, Eq.~\ref{7}.
We want to find an operator $K$, i.e., a matrix $K$ that anticommutes with $\mathcal{H}$ and, such that it satisfies $K.K=1$. Imposing these conditions, $\{K,\mathcal{H}\}=0$ and $K^2=1$, we find they can be satisfied with $K=\sigma_{\rm x} \otimes \sigma_0$.  Applying the same unitary transformation $UKU^T$ that diagonalizes the matrix $K$, to the Hamiltonian $\mathcal{H}$, we obtain,
$\mathcal{H}^{\prime}=U\mathcal{H}U^T$ as,
$$\mathcal{H}^{\prime}=
	\begin{bmatrix} 
	0 & A \\
	A^* & 0 \\
	\end{bmatrix}
	\quad
	$$
where,
$$A=
	\begin{bmatrix} 
	-2( \epsilon_k+h-\mu) & -2(\Delta_k+\lambda_k) \\
	-2(\Delta_k-\lambda_k)  & -2( \epsilon_k-h-\mu)  \\
	\end{bmatrix}
	\quad
	$$
with $\epsilon_{k} = -2 t \cos k $,  $\Delta_k=  2 i \Delta \sin k$ and $\lambda_k= 2 i \lambda \sin k$. This chiral symmetry of the Hamiltonian is due to the particle-hole symmetry of the spectrum that we observe for all values of the parameters, even in the presence of a magnetic field. 

In order to calculate the topological invariant we first obtain,
\begin{equation}
G(k)=\frac{\partial \ln \det[A(k)]}{\partial k},
\end{equation}
which is given by,
\begin{equation}
G(k)=\frac{-4t \mu \sin k -4\left(t^2 - (\Delta^2 - \lambda^2)\right) \sin 2 k}{-h^2+\mu^2 + 4 t \cos k (\mu+t \cos k) +4(\Delta^2 - \lambda^2) \sin^2 k}.
\end{equation}
Introducing,
\begin{equation}
\label{inva}
M(k)=\frac{1}{2\pi i}\int G(k) dk
\end{equation}
and performing the integration, we get
\begin{equation}
M(k)=\frac{1}{2 \pi} \Im m \left[ \ln(-h^2+\mu^2 + 4 t \cos k (\mu+t \cos k) +4(\Delta^2 - \lambda^2) \sin^2 k)\right].
\end{equation}
The topological invariant  is obtained from,
\begin{equation}
 \mathcal{M}=2\left[M(\pi)-M(0)\right].
\end{equation}
We get,
\begin{itemize}
\item
$\mathcal{M}=1$, for $(2t-\mu)^2 < h^2$ and $(2t+\mu)^2 >h^2$.
\item
$\mathcal{M}=-1$, for $(2t+\mu)^2 < h^2$ and $(2t-\mu)^2 >h^2$
\item
$\mathcal{M}=2$, for $h=0$ and $\mu<2t$.
\item
$\mathcal{M}=0$, for $h=0$ and $\mu>2t$.
\end{itemize}
For $\lambda=h=0$ and $|\mu/2t|<1$ we find $\mathcal{M}=2$, due to the two pairs of Majoranas in the uncoupled chains, as we obtained  previously,  and in agreement  with simulations.
\begin{figure}[ht]
\centering
\includegraphics[width=0.45\columnwidth]{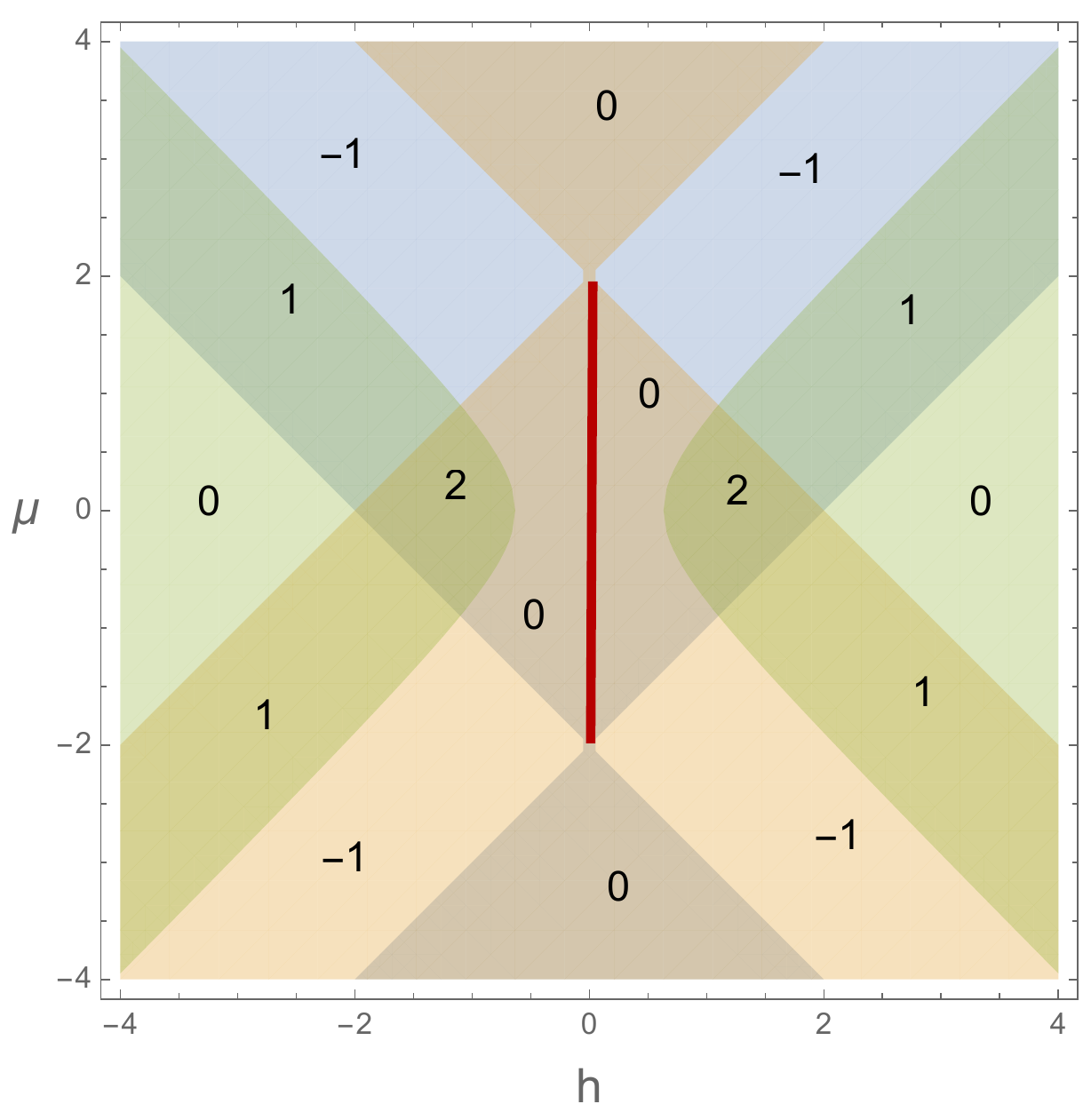}
\caption{(Color online) Topological indexes $\mathcal{M}$ of the different regions of the phase diagram. The red line, corresponding to $h=0$, $|\mu/2t|<1$,  has $\mathcal{M}=2$. To obtain the gap-closing lines at ($\pi/2)$, we used $\Delta=0.5$ and $\lambda=0.1$, in units of $t$.   Phases with topological indexed (1,-1) and (2,-2) are gapless phases with one and two pairs of Fermi points, respectively.}
\label{fig7}
\end{figure}

Since the gap may also close at $k=\pi/2$, we have additionally~\cite{lawz,law}, 
\begin{equation}
 \mathcal{M}=2 M(\pi/2)-M(\pi)-M(0)
\end{equation}
that yields,
\begin{itemize}
\item
$\mathcal{M}=2$, for $(2t \pm \mu)^2 > h^2$ and $h^2 > 4(\Delta^2-\lambda^2)+\mu^2$.
\item
$\mathcal{M}=1$, for $h^2> \mu^2+4(\Delta^2-\lambda^2)$ and \\
\subitem $(2t + \mu)^2 > h^2>(2t - \mu)^2$ or
\subitem $(2t - \mu)^2 > h^2>(2t + \mu)^2$
\item 
$\mathcal{M}=0$,  for $h^2> \mu^2+4(\Delta^2-\lambda^2)$ and $h^2< (\mu \pm 2t)^2$.
\end{itemize}
In these cases the indexes are not universal in the sense that the phase boundaries depend on the parameters $\Delta$ and $\lambda$.  The latter competes with superconductivity and it's role is to renormalize (reduce) the superconducting gap.  The topological indexes of the different phases are shown in Fig.~\ref{fig7}.

It is interesting to look at the phase diagram in Fig.~\ref{fig7}  together with  the dispersion relations shown in Fig.~\ref{fig3}b. These relations  were obtained for  $\mu=0$ and $2t>h>\sqrt{4(\Delta^2-\lambda^2)}$. This region of the phase diagram is  associated with a topological index $\mathcal{M}=2$ that in this case counts the number of pairs of topological Fermi points in this phase. The topological charge of a Fermi point is characterized by a winding number, similar to the one we calculated. It is integrated in a closed contour in momentum space that embraces the Fermi point~\cite{volovik,3dSOC}. In a one-dimensional system, this involves integrating all along the Brillouin zone,  as  in the calculation of the winding number, Eq.~\ref{inva}. 

\begin{figure}[ht]
\centering
\includegraphics[width=0.45\columnwidth]{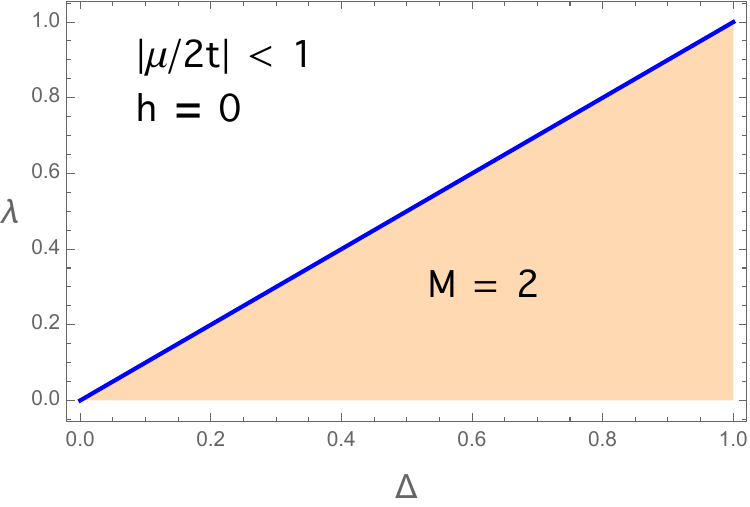}
\caption{(Color online) Region of the phase diagram, in color, where protected Majorana edge modes are observed. It is characterized by a topological index $\mathcal{M}=2$, and requires that $h=0$ and $|\mu/2t|<1$. The uncolored region, $\lambda > \Delta$,   has no Majorana edge modes.}
\label{fig8}
\end{figure}

It is also relevant that the region of the phase diagram with $\mu=0$ and $h<\Delta$, where we observed
magnetically polarized edge modes, is topologically trivial, with $\mathcal{M}=0$. This implies that these modes are not symmetry protected.

Finally,   Fig.\ref{fig8}  shows the region of the phase diagram where protected  Majorana modes are observed. They occur only at $h=0$, for $|\mu|/2t|<1$ and $\lambda < \Delta$.

\section{Conclusions}

In this work we presented a study of the topological properties of a superconducting chain with  electrons pairing in a spin triplet state $S=1$,  but with the zth-component $S^z=0$. The motivation for this study is that this type of pairing can be  induced in a chain whose orbitals have an anti-symmetric hybridization with those of a BCS $s$-wave superconducting substrate.  Alternative pairings to the simple spinless case studied by Kitaev have been considered to obtain  topological superconductors. In particular,  equal spin pairing \cite{tewari} and $d$-wave pairing~\cite{lawz,law}  were proposed. Here we studied  the case $S=1$, $S^z=0$  with a concrete physical motivation. We have obtained the symmetry properties and calculated the topological indexes  of the model and obtained a rich phase diagram with trivial and topological phases distinguished by these indexes. 

We pointed out the existence of a topological phase, in the absence of a magnetic field, with four Majorana modes, two in each edge of the finite chain. In small finite fields  $h < \Delta$, these modes become spin polarized. Each pair of edge modes  gives rise to a full electronic spin that can align parallel or antiparallel to the magnetic field. A full polarized electronic spin has a wave-function on both edges of the chain and corresponds to a  delocalized quasi-particle.     These edge modes are not symmetry protected, as they occur in a region of the phase diagram characterized by a trivial topological index, $\mathcal{M}=0$. 
For magnetic fields $h >\Delta$ these edge modes disappear as they merge with the spectrum of bogoliubons. This occurs before the gap closes at $h=2 \Delta$. 

For $h>2\sqrt{\Delta^2-\lambda^2}$ the topological excitations are in momentum space  and correspond to pairs of Fermi points.   At these points the dispersion relations are linear in momentum. These Fermi points   occur in a large region of parameters of the phase diagram.  They appear in pairs (Weyl fermions) and can only be destroyed by annihilating each other~\cite{3dSOC,volovik} at the time-reversal wave-vectors of the Brillouin zone. We have found nontrivial phases~\cite{nature} with a single pair, $\mathcal{M}=1$ and two pairs of Fermi points, $\mathcal{M}=2$.

\acknowledgements E. Silva would like to thank CAPES for financial support. R.C.B.R would like to thank FAPERJ and  CAPES, for financial support.   H.C. would like to thank FAPEMIG for partial financial support. H.C. also thanks the Centro Brasileiro de Pesquisas F\'isicas-CBPF, where part of this work was done, for the kind hospitality. M.A.C would like to thank CNPq and FAPERJ for partial financial support and Tobias Micklitz and Wei Chen for useful discussions.


\begin{thebibliography}{41}
\bibitem{note}  Let us write Eq.~\ref{eq1} as, $\Delta_{ind}^p=g(k) \Delta_a^s$, where $\Delta_a^s=(\Delta_k^{\uparrow \downarrow}+\Delta_k^{\downarrow \uparrow})/2$ is the symmetrized BCS pairing of the $a$-band (substrate), with $\Delta_k^{\sigma -\sigma}=\langle a^{\dagger}_{k \sigma}a^{\dagger}_{-k -\sigma}\rangle$. We then obtain for the induced pairing that, if $g(k)$ is symmetric,   $\Delta_{ind}^p=0$, and if $g(k)$ is antisymmetric, i.e., $g(-k)=-g(k)$, then, $\Delta_{ind}^p= g(k)  \langle a^{\dagger}_{k \uparrow}a^{\dagger}_{-k \downarrow}\rangle$.
\end{thebibliography}
\end{document}